\documentstyle[12pt,epsf]{article}

\newcommand{\nc}{\newcommand}
\nc{\beq}{\begin{equation}}
\nc{\eeq}{\end{equation}}
\nc{\beqa}{\begin{eqnarray}}
\nc{\eeqa}{\end{eqnarray}}

\input epsf
\newwrite\ffile\global\newcount\figno \global\figno=1

\def\writedef#1{}
\def\figin{\epsfcheck\figin}\def\figins{\epsfcheck\figins}
\def\epsfcheck{\ifx\epsfbox\UnDeFiNeD
\message{(NO epsf.tex, FIGURES WILL BE IGNORED)}
\gdef\figin##1{\vskip2in}\gdef\figins##1{\hskip.5in}
\else\message{(FIGURES WILL BE INCLUDED)}%
\gdef\figin##1{##1}\gdef\figins##1{##1}\fi}
\def\figinsert{}
\def\ifig#1#2#3{\xdef#1{fig.~\the\figno}
\writedef{#1\leftbracket fig.\noexpand~\the\figno}%
\figinsert\figin{\centerline{#3}}\medskip\centerline{\vbox{\baselineskip12pt
\advance\hsize by -1truein\center\footnotesize{  Fig.~\the\figno.} #2}}
\bigskip\endinsert\global\advance\figno by1}
\def\endinsert{}

\begin{document}



\title{\large{\bf Global Spread of Infectious Diseases}}

\author{
S.~ Hsu$^1$\thanks{hsu@duende.uoregon.edu} ~$~ \& ~$
A.~ Zee$^2$\thanks{zee@itp.ucsb.edu}\\
\\
$^1$Institute of Theoretical Science, \\
University of Oregon, \\
Eugene, Oregon 97403-5203 \\ \\
$^2$Kavli Institute for Theoretical Physics, \\
University of California, \\
Santa Barbara, California 93106-4030\\
\\}

\maketitle

\vspace{-24pt}

\begin{abstract}
We develop simple models for the global spread of infectious
diseases, emphasizing human mobility via air travel and the
variation of public health infrastructure from region to region.
We derive formulas relating the total and peak number of
infections in two countries to the rate of travel between them and
their respective epidemiological parameters.
\end{abstract}


\newpage

\section{Introduction and linear model}

Recent outbreaks of atypical pneumonia (SARS) \cite{SARS} as well
as likely future epidemics and
pandemics of influenza \cite{Influenza} provide
ample motivation for the study of the global propagation of
infectious diseases. (For an overview of modelling of infectious
diseases in humans, see \cite{AM}). In this letter we construct a
class of simple models of this phenomena, with particular
attention to human mobility and the variation of public health
capabilities across national and regional boundaries. Spatial dynamics
involving the diffusion equation,
of the type studied in \cite{Murray} (as in, e.g., the spread of
rabies in fox populations), is of less interest to us than regional
or national variation of parameters. We treat each region as spatially
homogenous and focus on mobility (e.g., via air travel) as the mechanism
by which disease is transmitted.

The population variables in our model: $S$ = susceptibles, $I$ =
infected, $R$ = recovered and $D$ = deceased, are functions of time $t$
and of {\it discrete} geographical location labelled by an index
$i=1,\cdots,N$. Thus our equations describe the time evolution of a $4N$
dimensional vector:
\begin{equation}
\vec{x} = \left( \begin{array}{c} S_1 \\ I_1 \\ R_1 \\ D_1 \\ S_2
\\ I_2 \\ . \\ . \\ . \end{array} \right)
\end{equation}

In the simplest linear model, the dynamics of $\vec{x}$ are
characterized by a $4N~\times~4N$ matrix whose entries are
probabilities per unit time: $t_i$ = probability of transmission
of disease from an infected to susceptible individual in region
$i$, $r_i$ = probability of recovery of infected individual in
region $i$, $d_i$ = probability of death of infected individual in
region $i$, $m_{i \rightarrow j}$ = probability of movement of an
{\it infected} individual from country $i$ to country $j$. Each of
these probabilities incorporates several additional parameters
that would appear in a more complicated model. For example, $t_i$
depends on other factors such as the probability and effectiveness
of quarantine, the length of time an infected individual remains
asymptomatic, the population density, etc. A very simple
modification would be to include asymptomatic as well as
symptomatic carriers of the infection by expanding our matrix to
$5N~\times~5N$ dimensions. Contagious asymptomatics would have a
large transmission probability, as they are difficult to
quarantine. We will return later to enhancements of the basic
model.

The basic parameters $r_i ,t_i ,d_i ,m_{i \rightarrow j}$ vary
from region to region, and can be estimated from epidemiological
data. Our organizing principle for dividing the world into regions
is the relative homogeneity of these parameters within each
region. This division may or may not follow national boundaries,
as the parameters may vary significantly (e.g., from urban to
rural regions) within a given country.

The {\it basic reproductive rate} \cite{AM}, commonly
denoted ${\cal R}_0$, is defined as the average number of secondary
cases caused by a single infected individual. A simple computation
(neglecting migration) for region $i$ yields
\begin{equation}
{\cal R}_0^i ~=~ t_i ~+~ (1 - r_i - d_i)t_i ~+~  (1 - r_i - d_i)^2 t_i
~+~ \cdots ~=~ \left( {t_i \over r_i + d_i } \right)~~~.
\end{equation}
When ${\cal R}_0 > 1$ the infected population grows exponentially.

The system of equations governing the time evolution of $\vec{x}$
is
\begin{equation}
\label{ME} {d \vec{x} \over dt} = M \vec{x} + \vec{J}~~~,
\end{equation}
where $\vec{J}$ is a possible source of infected individuals
coming from zoonosis, an animal reservoir of disease. All elements
of $\vec{J}$ are zero except the $(4i-2)$ and $(4i-3)$ entries
corresponding to an animal reservoir in region $i$ which
contributes to $\, d I_i /  dt \,$ and $\, - \, d S_i / dt \,$.
Note that, unlike the well-known SIR model \cite{Murray}, equation
(\ref{ME}) is linear. The transmission rate per unit time is given
by $t_i I_i$ rather than $t_i I_i S_i$. This makes our model
slightly less realistic, as the transmission rate per infected
individual does not drop as the number of susceptibles  goes to
zero. The SIR model predicts some fraction of uninfected
susceptibles $S( t \rightarrow \infty )$ even when ${\cal R}_0 >
1$, whereas our model would predict that each individual in the
entire population is eventually be infected. This leads to an
overestimate relative to the SIR model of the total number of
infected in countries with ${\cal R}_0 > 1$. However, the
difference is only numerically significant when ${\cal R}_0$ is
close to 1. (We do not expect the input parameters to be
sufficiently well-determined that, e.g., a factor of two in the
predicted number of infections is very significant.) Another
method for cutting off the exponential growth is to add
(nonlinear) logistic terms of the form $- k_i I_i^2$ to the right
hand side of the equation, at the cost of adding additional
parameters.

Equation (\ref{ME}) can be solved analytically:
\begin{equation}
\vec{x} = - M^{-1} J ~+~ e^{Mt} \left( \vec{x}_0 + M^{-1} J
\right)~~~, \label{SME}
\end{equation}
where $\vec{x}_0$ is the initial condition. The importance of the
variables $S_i,R_i,D_i$ is to impose the overall conservation of
humans, which is a consequence of the form of $M$, and the
constraint that $S,I,R,D$ be positive semi-definite at all times.
Although equations (\ref{ME}) and (\ref{SME}) appear simple, the
constraints lead to a nontrivial system.

Below we exhibit the matrix M for the case of two countries,
$i=1,2$.
\begin{equation}
\label{ansatz1} M  ~=~ \left(\begin{array}{cccccccc}
0 & -t_1 & 0 & 0 &0&0&0&0\\
0 & -r_1 - d_1 - m_{1 \rightarrow 2} + t_1 & 0 &0 &0&m_{2 \rightarrow 1}&0&0 \\
0 & r_1 & 0 & 0 &0&0&0&0 \\
0& d_1 &0&0 &0&0&0&0 \\

0 & 0 & 0 & 0 &0& -t_2 &0&0\\
0 & m_{1 \rightarrow 2} & 0 &0 &0& -r_2 - d_2 - m_{2 \rightarrow 1} + t_2 &0&0 \\
0 & 0 & 0 & 0 &0& r_2&0&0 \\
0& 0 &0&0 &0& d_2&0&0 \\
\end{array} \right)~~~.
\end{equation}
Note the sum rule implicit in this matrix: $\sum_i^{4N} x_i = {\rm
constant}$, which reflects conservation of humans.

Despite the simplicity of these models, they may be useful tools
for public health policy. For example, the economic impacts on
developed country economies from diseases in developing
nations can be estimated. This data can be used in cost-benefit
assessments of improvements in public health infrastructure in
both developed and developing nations.

\section{Linear model: analytics and simulations}

We now discuss some of the analytic properties of our linear
models. First, we note that the system of $4N$ equations can be
reduced to a smaller number of equations, due to the large number
of zeroes in $M$. In essence, the dynamics involves only the
infected individuals $I_i$: the evolution of the $S_i,R_i$ and
$D_i$ variables are dependent on $I_i$. Thus there are really only
$N$ equations (but subject to constraints; see below), as can be
seen from the fact that $M$ has only $N$ non-zero eigenvalues. The
matrix $M$, restricted to the $N \times N$ subspace of infected
populations $I_i$, is diagonal up to the (small) off-diagonal
mobility entries $m_{i \rightarrow j}~$. If the mobility matrix
were symmetric, we could diagonalize $M$ via an orthogonal change
of basis. However, in general $m_{i \rightarrow j} \neq m_{j
\rightarrow i}~$, so we need to use a bi-linear transformation.
That is, we can obtain a diagonal matrix $M_D$
\begin{equation}
M_D ~=~ L \, M \, R~~~,
\end{equation}
where $L = R^{-1} = R^{T}$ up to corrections which vanish for
symmetric $m_{i \rightarrow j}$. In the basis $\vec{y} = R^{-1}
\vec{x}$, equation (\ref{ME}) becomes
\begin{equation}
L \, R \, {d\vec{y} \over dt} ~=~ M_D \, \vec{y} ~+~ L \vec{J}
~~~.
\label{MEy}
\end{equation}
While the eigenvalues $\lambda_i$ tell us a great
deal - essentially, what would happen in each region if it were
isolated - there is additional information in the $L,R$ matrices
which relate the original regional basis to the diagonal basis.
Below we investigate explicit scenarios involving a developing
country  $i$ with positive eigenvalue $\lambda_i$, and a
developed country $j$ with eigenvalue $\lambda_j$ which is
either negative, or, if positive, smaller than $\lambda_i$. In
these scenarios the $m_{i \rightarrow j}$ mobility is much more
important than the reverse migration $m_{j \rightarrow i}$, since
the spread of disease is mainly unidirectional. Hence, we could
simply assume $m_{i \rightarrow j} = m_{j \rightarrow i}$ without
changing the results appreciably. In that case, $L = R^{-1}$, so equation
(\ref{MEy}) becomes uncoupled.

In the $N = 2$ case, the non-zero eigenvalues are
\begin{eqnarray}
\lambda_1 &=& -r_1 - d_1 - m_{1 \rightarrow 2} + t_1 \nonumber \\
\lambda_2 &=& -r_2
-d_2 - m_{2 \rightarrow 1} + t_2~~~.
\end{eqnarray}
Note that a positive eigenvalue $\lambda_i$
corresponds to ${\cal R}_0^i > 1$.

Two generic implications can be stated, depending on the sign of
$\lambda_i$. Here we assume that the entries of the off-diagonal
mobility matrix $m_{i \rightarrow j}$ are small compared to the
other probabilities, so that we can discuss the eigenvalue
associated an individual region $i$. This is likely to be the case
in any realistic scenario where the mobility matrix reflects air
travel.

A country with positive eigenvalue $\lambda_i$ will experience
exponential growth in the number of infecteds $I_i$. This growth
is only limited by the constraint of total population: eventually
roughly the entire susceptible population has been infected and
after a long
period of time there are only recovered and dead individuals.
(This situation is familiar in the case of the common cold or
influenza.) There are two important timescales involved:
$\lambda_i^{-1}$, which determines the rate of exponential growth,
and $|\lambda_i - t_i|^{-1}$, ($\lambda_i - t_i$ is negative)
which gives the rate of decline of $I_i$ after saturation occurs
and there are no more susceptibles to be infected ($S_i = 0$, so
effectively the transmission probability $t_i$ can be set to zero
in the corresponding eigenvalue). The outbreak in a highly
susceptible country cannot last much longer than the sum of these
timescales.

A country with a negative eigenvalue is capable of suppressing any
outbreaks of the disease. Nevertheless, there may be a steady
state number of infections
\begin{equation} I_j ~\sim~ {J_j \over |\lambda_j|} ~+~
\sum_{i \neq j} {m_{i \rightarrow j}~ I_i \over | \lambda_j -
\lambda_i|}
\end{equation}
due to the animal reservoir $J$ and/or the migration of infected
individuals from other regions. This equation is valid for $m_{i
\rightarrow j}$ small compared to $|\lambda_j - \lambda_i|$. (The
difference in eigenvalues results from the diagonalization of $M$,
and is analogous to the result in quantum mechanics for the change
in energy eigenstate under a small perturbation to the
Hamiltonian.) The most dangerous sources of disease migration are
countries $i$ with positive eigenvalue $\lambda_i$. The total flux
of migration is roughly determined by the mobility $m_{i
\rightarrow j}$, the population $S_i (0)$ and the timescales
$\lambda_i^{-1}$ and $|\lambda_i - t_i |^{-1}$.

Consider a plausible worst case scenario involving a developing
country $i$ with positive $\lambda_i$ and much larger population
than developed country $j$ which has negative $\lambda_j$. The
developing country is the main source of infection for the
developed country. In this case, we obtain the simple relation
\begin{equation} I_j ~\sim~ {m_{i \rightarrow j}~ I_i \over |
\lambda_j - \lambda_i|}
\end{equation}
relating the number of infected in the two countries at any
particular time. The {\it total} number of infections in country
$j$ (during the entire epidemic) is related to $I_j (t)$ in a
complicated way, depending on recovery and death rates. However,
by solving equation (\ref{SME}) we can obtain a simple expression
for the ratio of total number of infections $N_i$ in the two
countries
\begin{equation}
\label{NR} {N_j \over N_i}  ~=~ {m_{i \rightarrow j} \over \vert
\lambda_j \vert} \, \left( { {d_j + r_j} \over {d_i + r_i}}
\right)~~~,
\end{equation}
to leading order in $m_{i \rightarrow j}$. The first ratio in this
result can be interpreted as the fraction of infected individuals
who travel from country $i$ to $j$ in the timescale $\vert
\lambda_j \vert^{-1}$, which is roughly the ``halving'' time in
country $j$. As long as the death and recovery rates in the two
countries are not too different, this can be used as a rule of
thumb to estimate the ratio of total number of infections.
Obviously it behooves country $j$ to make $\vert \lambda_j
\vert^{-1}$ as small as possible.

We can also obtain simple expressions for the peak number of
infected individuals in both countries (neglecting animal
reservoirs):
\begin{equation}
I_i^{\rm peak} ~=~ {\lambda_i \over t_i} \left( P_i - I_i(0)\left(
1 + \frac{t_i}{\lambda_i} \right) \right) ~ \simeq ~ {\lambda_i
\over t_i} P_i \label{peak1}
\end{equation}
and
\begin{eqnarray} \label{peak2}
I_j^{\rm peak} &=& {m_{i \rightarrow j} \over \vert \lambda_i -
\lambda_j \vert} \left( \frac{\lambda_i}{t_i} \, \left( P_i -
I_i(0)\left( 1 + \frac{t_i}{\lambda_i} \right) \right) ~+~ I_i(0)
\left(
{\lambda_i \over I_i(0) t_i}
\left( P_i - I_i(0)\left( 1 +
\frac{t_i}{\lambda_i} \right) \right)
\right)^{\lambda_j \over
\lambda_i} \right)  \nonumber \\
&\simeq& {m_{i \rightarrow j} \over \vert \lambda_i - \lambda_j
\vert} {\lambda_i \over t_i} P_i
\end{eqnarray}
where $I_i(0)$ is the initial number of infecteds, and $P_i$ is
the total population, in country $i$. This gives the maximum
capacity required of the medical system during the epidemic.
Again, our result is likely to be an overestimate relative to a
more realistic SIR or logistic model.

Below we describe the results of two illustrative simulations of
the linear model. The parameters are chosen to be realistic
average probabilities per week of transmission, recovery, death
and international travel. We integrate equation (\ref{ME}) forward
numerically, discretizing time in units of weeks, and starting
with a random number of infecteds between 1 and 10. Country 1 has
population $10^9$ and country 2 has population $10^8$. Note that
the figures display the infected population $I_i (t)$ as a
function of time, not the total number of individuals who have
ever been infected. The latter quantity, although not displayed,
agrees precisely with the result given in equation (\ref{NR}). The
peak infected results in equations (\ref{peak1}) and (\ref{peak2})
are also confirmed.

\begin{figure}[htb]
\center{ \leavevmode \epsfbox{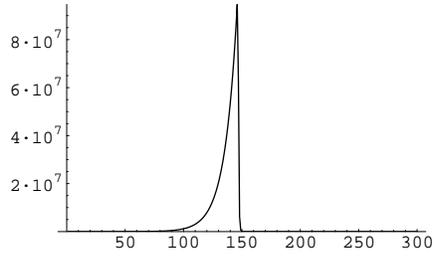} \caption{Number of
infected in country 1, simulation 1. Note saturation and cutoff.}
\label{C11}}
\end{figure}

1) One positive and one negative eigenvalue. The epidemic infects
the entire population of country 1 (figure (\ref{C11})), but
infects only a small fraction of the population of country 2
(figure (\ref{C21})). The parameter values used are $t_1 = 1;~ r_1
= 0.7;~ d_1 = 0.2;~ t_2 = 0.85;~ r_2 = 0.9;~ d_2 = 0.05;~ m_{12} =
0.00001;~ m_{21} = 0$, eigenvalues $\lambda_1 = 0.09999$,
$\lambda_2 = -0.1$. The overall results are consistent with our
result (\ref{NR}).

\begin{figure}[htb]
\center{ \leavevmode \epsfbox{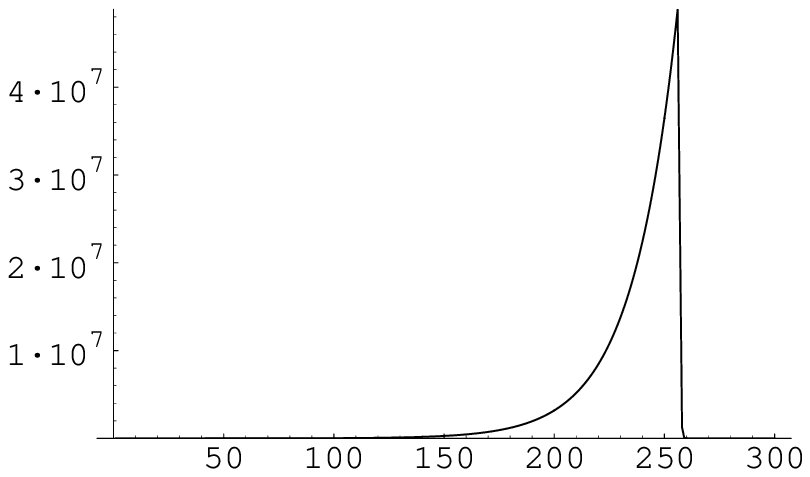} \caption{Number of
infected in country 1, simulation 2. Note saturation and cutoff.}
\label{C12} }
\end{figure}

\begin{figure}[htb]
\center{ \leavevmode \epsfbox{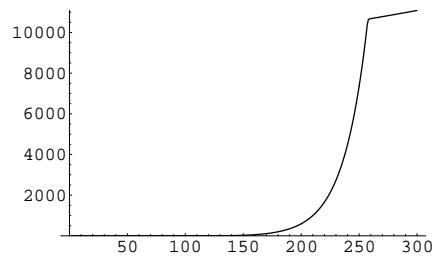} \caption{Number of
infected in country 2 (small positive eigenvalue), simulation 2.
Rapid growth driven by migration from country 1 levels off after
epidemic subsides in country 1. Subsequent growth in country 2 is
much slower.} \label{C22} }
\end{figure}

\begin{figure}[htb]
\center{ \leavevmode \epsfbox{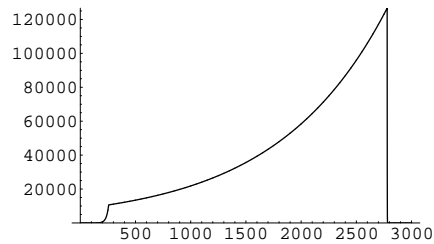} \caption{Number of
infected in country 2 (small positive eigenvalue), simulation 2.
This graph extends over a longer period of time, exhibiting
eventual saturation.} \label{C22e} }
\end{figure}

2) Two positive eigenvalues. The epidemic sweeping through country
1 (figure (\ref{C12})) drives rapid growth in country 2 (figures
(\ref{C22}),(\ref{C22e})) until it has completed its course. At
this point slower growth characterized by $\lambda_2$ (which is
very small) resumes. In the end all susceptible individuals in
both countries are infected. The falloff of $I_2$ appears abrupt
in figure (\ref{C22e}) because $|\lambda_2 - t_2|$ is much larger
than $\lambda_2$. $t_1 = 1;~ r_1 = 0.75;~ d_1 = 0.2;~ t_2 =
0.851;~ r_2 = 0.8;~ d_2 = 0.05;~ m_{12} = 0.00001;~ m_{21} = 0$,
eigenvalues $\lambda_1 = 0.04999$, $\lambda_2 = 0.001$.

The numerical values of parameters have been varied only slightly
between the two examples, but the resulting behaviors are very
different, largely because the eigenvalue $\lambda_2$ has changed sign.

\bigskip

\section{Nonlinear models}

To make the models more realistic, we can let the parameters
$t_i,r_i,d_i$ depend on local quantities $x_i$. The resulting
equations are nonlinear, although still straightforward to
simulate numerically. For example, it is possible that a region's
medical capabilities are overwhelmed as the number of infected
individuals increases. Indeed, in both the SARS outbreak and the
influenza of 1918 (Spanish flu) medical personnel were
disproportionately affected.

\begin{figure}[htb]
\center{ \leavevmode \epsfbox{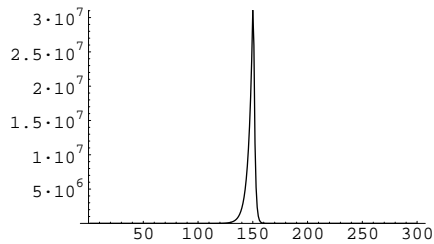} \caption{Number of
infected in country 2, simulation 3. Initially negative eigenvalue
becomes positive as number of infected increases, leading to
saturation in country 2.} \label{C23} }
\end{figure}

As a simple example, we take the parameters from simulation 1 in
the previous section, and let the recovery rate $r_2$ of country 2
(which had a negative eigenvalue) interpolate between the initial
value $r_2$ when $I_2 = 0$ and $r_2^*$ when $I_2$ is large,
according to the formula
\begin{equation}
r_2 (I_2) = r_2  +  (r_2^* - r_2) \left( {I_2 \over I_2 + I_2^*}
\right)~~~.
\end{equation}
This reflects a saturation of medical treatment capabilities after
some characteristic number of infections. When $I_2$ is much
larger than $I_2^*$ the recovery probability per unit time is
reduced from its initial value of $r_2$ to $r_2^*$, and the
eigenvalue $\lambda_2$ may become positive. We take $I_2^* =
2000$, $r_2 = .9$ and $r_2^* = .7$. The results are shown in
figures (\ref{C23}) and (\ref{C23a}). Figure (\ref{C23a}) reveals
the onset of the nonlinear behavior when $I_2$ is a few thousand
(compare to figure (\ref{C21})).

\section{Internet viruses and worms}

These models can be modified to describe the propagation of
computer viruses and worms. (From the point of view of our models,
there is no essential difference between the two.) A virus
commonly inserts itself into other program files, in the same
manner that a virus in nature takes over the workings of normal
cells.  When the infected program runs, the virus code gets a
chance to inspect its environment and look for and infect new
carriers in the form of other program files.  A worm
is a self-replicating program that does not
alter files but resides in active memory and duplicates itself by
means of computer networks. Worms use facilities of an operating
system that are meant to be automatic and invisible to the user.
It is common for worms to be noticed only when their uncontrolled
replication consumes system resources, slowing or halting other
tasks.

The variation of parameters by region in our model would be a
consequence of different levels of security administration in
home, small business, educational and corporate networks.

Mobility $m_{i \rightarrow j}$ has no analog here, as the infected
computer does not hop from network to network. Rather, an infected
machine can infect other machines connected to the Internet,
leading to a non-local (off-diagonal) transmission matrix $t_{i
\rightarrow j}$. Machines on local networks $k$ with strong
firewalls and virus scanning would be difficult to infect, leading
to correspondingly small entries $t_{i \rightarrow k}$. However,
once a virus or worm has penetrated the firewall, spread within a
local network can be quite rapid, so the diagonal elements $t_{k
\rightarrow k}$ may be quite large. Often, the only factor
limiting the rate of infection is saturation of available
bandwidth by worm scanning activity.

``Slammer'', the most successful worm to date, exploited a
security hole in Microsoft's SQL Server (a database server), and
infected 75,000 machines around the world in less than a half hour
on January 25, 2003 \cite{Slammer}. The initial doubling time of
the infected population was 8.5 seconds, and the growth curve
displayed a typical logistic shape.

\section{Discussion}

We have attempted to model the spread of infectious diseases in a
global environment, taking into account geographical variation of
public health capabilities as well as human mobility (i.e., via
air travel). We believe our models capture a large range of
possible behaviors using a minimal set of parameters, each of
which can be roughly estimated from data.

\begin{figure}[htb]
\center{ \leavevmode \epsfbox{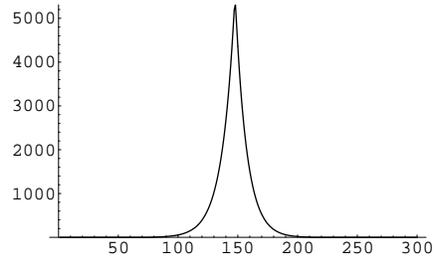} \caption{Number of
infected in country 2 (negative eigenvalue), simulation 1.}
\label{C21}}
\end{figure}

\begin{figure}[htb]
\center{ \leavevmode \epsfbox{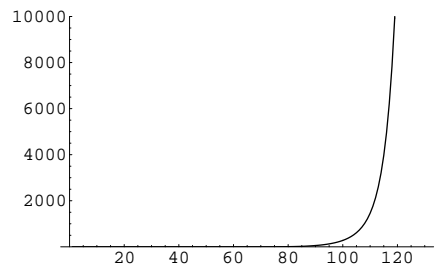} \caption{Number of
infected in country 2, simulation 3. Detail of non-linear region
in which recovery probability per week falls from .9 to .7.
Compare to figure (\ref{C21}).} \label{C23a} }
\end{figure}

One interesting conclusion from our models is that typical
international mobility -- the probability per unit time of
international travel for a given infected individual, estimated at
$m_{i \rightarrow j} \sim 10^{-5}$ per week -- is still
sufficiently small that a country with well-developed public
health infrastructure (effectively, a negative eigenvalue
$\lambda$) can resist an epidemic even when other more populous
countries experience complete saturation. In the quasi-realistic
simulation 1 (figures (\ref{C11}),(\ref{C21})), of order $10^5$
infections occur in country 2, even though the disease has swept
completely through country 1. In reaching this conclusion, we kept
the mobility parameter fixed during the outbreak, and did not
assume any draconian quarantine on international travellers
arriving in country 2. Such measures would reduce the number of
infections in country 2 considerably. Of course, this conclusion
assumes that the public health infrastructure in country 2 remains
robust during the outbreak. In the nonlinear simulation 3 (figures
(\ref{C23}), (\ref{C23a})), we see that a breakdown in the medical
system can lead to grave consequences.

In the case of two countries, one of which is a ``reservoir'' with
positive eigenvalue $\lambda_i$ and the other with negative
eigenvalue $\lambda_j$, a good rule of thumb arising from equation
(\ref{NR}) is that the ratio of total number of infections in the
two countries is given by the fraction of infected individuals who
migrate in a timescale $\vert \lambda_j \vert^{-1}$, which is the
``halving'' time for the epidemic in country $j$. In our
simulation 1, this timescale is about two months, and the
fractional mobility over that period is $\sim 10^{-4}$, leading to
$10^5$ infections in country 2 if the entire $10^9$ population of
country 1 is infected. The maximum number of infections at any
given time in the simulation is $5 \cdot 10^3$ (figure
(\ref{C21})). If the medical system of country 2 can treat this
number of patients without breaking down (entering the nonlinear
regime), it can prevent a larger outbreak.

Our formula (\ref{NR}) is not applicable to early SARS data since
the eigenvalues have clearly been time-dependent during the early
stages of the epidemic (estimates of ${\cal R}_0$ vary widely
during different time periods \cite{SARS}). Even developed
countries like Canada and Taiwan exhibited ${\cal R}_0 > 1$ during
the first months (positive eigenvalues), although as of this
writing eigenvalues appear to be negative for all countries.

\bigskip

\section*{Acknowledgements}
\noindent

AZ thanks A. Madhav for conversations.
SH is supported under Department of Energy
contract DE-FG06-85ER40224. AZ is grateful for the
warm hospitality of the University of Oregon, where this work
was begun, and
acknowledges support from NSF grant PHY 99-07949.


\bigskip


\newpage

%

\epsfysize=3.5 cm

\epsfysize=3.5 cm

\epsfysize=3.5 cm

\epsfysize=3.5 cm

\epsfysize=3.5 cm

\epsfysize=3.5 cm

\epsfysize=3.5 cm

\end{document}